\newif\iffiginc
\figinctrue   % use figures
\iffiginc
\documentstyle[prd,aps,preprint,tighten,floats,psfig]{revtex}
\else
\documentstyle[prd,aps,preprint,tighten,floats]{revtex}
\fi

\begin{document}
\preprint{\bf UG-FT-45/94\hskip0.5cm UPR--636--T}
\date{\bf January 1995}

\title{Reconstruction of the Extended Gauge Structure from $Z'$ Observables at
Future Colliders}

\author{F. DEL AGUILA}
\address{Departamento de F\'\i sica Te\'orica y del Cosmos,
Universidad de Granada \\
Granada, 18071, Spain}

\author{M. CVETI\v C and P. LANGACKER}
\address{Department of Physics, University of Pennsylvania \\
Philadelphia, PA 19104-6396}

\maketitle

\begin{abstract}
The discovery of a new neutral gauge boson $Z'$ with a mass in the
TeV region would allow for  determination of
gauge couplings of the $Z'$ to
ordinary quarks and leptons in  a model independent
way. We show that these couplings in turn would allow us  to determine
the nature of the extended gauge structure.
As a prime example we study the $E_6$ group. In this case
 two discrete constraints on  experimentally determined
  couplings have to be satisfied.  If so,
the couplings  would then uniquely determine
the two parameters, $\tan \beta$ and $\delta$,
which fully  specify the nature  of the $Z'$ within $E_6$.
If the $Z'$ is part of the $E_6$ gauge structure, then  for $M_{Z'}=1$ TeV
$\tan \beta$ and
$\delta$ could be determined to around $10\%$  at the future colliders.
The NLC provides a unique determination of  the two
constraints  as well as of $\tan \beta$ and $\delta$, though with
slightly  larger error bars than at the LHC. On the other hand,
since the LHC primarily determines  three out of four normalized couplings,
it  provides weaker constraints for the underlying gauge structure.
\end{abstract}
\pacs{12.15Cc, 13.38.+c, 13.85.Qk, 14.80.Er}

\section{Introduction}
\def\lsim{\lower3pt\hbox{$\buildrel<\over\sim$}}
\def\overtext#1{$\overline{#1}$}
\def\etal{{\it et al.}}
\def\to{\rightarrow}

New neutral ($Z'$) and charged ($W'$)  gauge bosons are a generic prediction of
theories beyond the standard
model. If their mass turns out to be in the
TeV region future hadron  colliders (the Large Hadron Collider (LHC) at CERN)
as well as $e^+e^-$  colliders  (the New Large Collider (NLC))  would provide
an ideal environment for their discovery as well as  for
further testing of their properties.

In  recent years  it has been demonstrated
that for  $M_{Z',W'}\sim 1$
TeV  a diagnostic study of heavy gauge bosons is possible  at  the
LHC (integrated luminosity ${\cal L}_{int}= 100$ fb$^{-1}$, center of mass
energy $\sqrt s=14$ TeV)\cite{ACL,ACLR}\footnote{Production rates
have recently been
recalculated in Ref. \cite{G}.} as well
as at
the NLC (${\cal L}_{int}=20$ fb$^{-1}$, $\sqrt s= 500$ GeV) \cite{DLRSV}.
Both machines turn out to  have  complementary diagnostic power for
$Z'$ physics \cite{AC}.

For $M_{Z'}\le 1-2$ TeV the future
colliders allow for model independent determination of gauge couplings to
quarks and leptons. The analysis assumes  family universality;
 $[Q',T_i]=0$, where $Q'$ is the $U(1)'$ charge and $T_i$
 are the generators of
$SU(2)_L$; and that $Z-Z'$ mixing is negligible.
The LHC can  probe
 the absolute magnitude of the
overall strength of the new interaction
$g_2$, as well as the magnitude of
(primarily) three out of four normalized gauge couplings.
 On the other hand,  the NLC with longitudinal    electron beam
 polarization and heavy  flavor tagging available
is  sensitive to the ratio of $g_2^2/M_{Z'}^2$ and
four normalized gauge couplings. The latter can also
 be determined
 in terms of normalized couplings probed directly at the  LHC;
the  error bars are
typically smaller by a factor of $\sim 2$   than those at  the NLC \cite{AC},
but there is a few-fold ambiguity.
Thus,  the LHC and the NLC are complementary and together
have the  potential to uniquely determine
the couplings with small error-bars. For $M_{Z'}\sim 1$ TeV the
error bars are in the 10\%-20\% region.

In this note  we point out that a model independent determination of gauge
couplings  in turn provides a tool to
learn more about the  nature of
the extended gauge structure.  In particular, one would be able to gain
information about the  underlying gauge group and subsequently probe
the type of symmetry breaking
 associated with the new heavy gauge boson.

To illustrate the method  we study as a prime example the
Cartan subalgebra of the $E_6$ gauge group, which
provides a general enough framework for grand unified symmetry
 with canonical embedding of electric charge and simple cancellation of
anomalies. In addition, it is  motivated by string theory. $E_6$
imposes two  discrete constraints  on the  $Z'$ gauge couplings.
Provided these are  satisfied by the  measured couplings,
one  can further  determine the two parameters $\tan \beta$
and $\delta$. These two parameters, together with the
overall strength of the new gauge couplings $g_2$, contain
the information brought by the new interaction concerning
the pattern of symmetry breaking, possible intermediate
mass scales, and renormalization effects above the $Z'$.
This window to higher scales, as in the case of the weak
angle, is not sufficient to fix the full theory at the
Planck scale, but it should allow for constraining
possible unification schemes, excluding many possibilities.
See, for instance, Refs.  \cite{AGQ,A} for a discussion of
the predictions for the new parameters of extended
models from the heterotic string.

In  Section II we spell out the formalism  and  parameterization of the $Z'$
interaction with the  corresponding  quark and lepton
neutral currents. In Section III we recapitulate
the results for the
 model independent determination  of gauge couplings.  In Section IV we
 illustrate how  such  couplings
allow for a determination of the underlying gauge symmetry and
subsequently for  determination of the symmetry breaking pattern with
a  particular gauge structure. Conclusions are given in Section V.

\section{Formalism}

The coupling of an additional neutral gauge boson $Z'$ and
the neutral standard model bosons
$W_3, B$ to ordinary fermions are parameterized
in the neutral current Lagrangian as
\cite{A,AMP}:
\begin{equation}
-{\cal L}_{NC}=
\bar{\psi}_k \gamma ^{\mu} [
T_{3k}g W_{3\mu} + Y_kg_{1} B_{\mu}
+ (Q'_kg_{2} + Y_kg_{12}) Z_{\mu}'
] \psi _k,
\label{lag}
\end{equation}
where $T_{3}$ and $Y$ are the standard model isospin and
hypercharge charges, respectively,  and $Q'$ is the charge of
the extra $U(1)'$. A sum over all the
quarks and leptons $\psi_k$ is implied.
This Lagrangian describes the general coupling of a new gauge
boson $Z'$ with charge $Q'$ which  commutes with the $SU(2)_L$ generators
$T_i$, {\it i.e.},  $[Q', T_i] = 0$. This is  the case for all of the
extended gauge structures for which the generator of the $U(1)'$
associated with $Z'$ lies in its Cartan subalgebra.
In the following we also assume family universality and
neglect $Z-Z'$ mixing,  as indicated by present bounds\cite{bounds} and the
expected accuracy at future colliders.

The $Z'$ couplings are specified by the (normalized) charges
$Q'_k$ and the overall strength $g_2$, as well as by $g_{12}$,
which is the  coupling of $Z'$ to the weak hypercharge.
Thus, the neutral current Lagrangian of the ordinary fermions,
which includes an additional gauge boson, is specified by five charges
$g_2Q_k'$
and the relative strength of the $Z'$ gauge coupling to the hypercharge $Y$
and the $Q'$ currents,  $\delta \equiv g_{12}/g_2$.
The charges $Q'_k$ are characteristic of the underlying
extended gauge structure.  $Q'_k$
charges for typical models based on  $SO(10)$ and $E_6$
 grand unified symmetries  are given in Table \ref{charges}.
The  $SO(10)$ symmetry  includes the general {\it left-right}
symmetric models with \cite{A}
\begin{equation}
g_2 = \frac {e}{c_W}\sqrt {\frac {2}{5}}
\frac {\rho ^2+1}{\rho},\ \delta = \frac {g_{12}}{g_2} =
\sqrt {\frac {1}{6}}\frac {3\rho ^2-2}
{\rho ^2+1},\ \rho = \sqrt {\kappa ^2
\cot^2\theta _W-1},
\label{lrm}
\end{equation}
where $\kappa = \frac {g_R}{g_L}$ is the ratio of the gauge couplings $g_{L,R}$
for  $SU(2)_{L,R}$, respectively.  In general,
  $\kappa > \frac {s_W}{c_W}$ \cite{CKL}.
Here  $c_W\equiv \cos\theta_W$ and $s_W\equiv \sin\theta_W$ where $\theta_W$
is the weak angle.

For a  particular
extended gauge symmetry the charges  $Q'_k$ are fixed
numbers which usually satisfy specific constraints.  For  $SO(10)$ the  $Q'_k$
satisfy  the four conditions:
\begin{equation}
Q'_{e^c_L} = Q'_{u^c_L} = Q'_{q_L},\ \ \
Q'_{\ell_L} = Q'_{d^c_L},\ \  \ Q'_{d^c_L} =
- 3\ Q'_{q_L}\ \ ,
\label{const}
\end{equation}
whereas  for $E_6$ only the first three  are satisfied in general.

The gauge couplings, $g, g_{1,2,12}$, specify the overall strength of
the interactions. They are fixed at the large scale, where the
full underlying gauge structure is present, {\it e.g.},
in string theory  the gauge couplings of all the group factors are related
at the string compactification scale.
However, these couplings  are renormalized  at low energies, where their values
  crucially depend on the matter content, the
symmetry breaking  pattern,
and the threshold effects. In particular,  $g_{12}$ may vanish
at the large  scale, but it is in general  non-zero   at low energy.
It allows for description of the
{\it left-right} models (see Eq. (\ref{lrm})) within the $SO(10)$ symmetry.
Introduction of $g_{12}$
is also necessary for a complete parameterization
of low energy models when only the
extended gauge structure or its Cartan subalgebra, within which
 $Q'$ lies, is
known\footnote{Without loss of generality
 the  gauge coupling  $g_{21}$, which  parameterizes the strength of the
$B$ coupling  to the neutral currents  specified by $Q'$, can be set to zero.
 Gauge  couplings of $B$ and $Z'$ to the neutral currents
 specified by  $T_3$ vanish by gauge invariance
\cite {AMP}.}.

\begin{center}
\begin{table}
\caption{Isospin $T_3$, hypercharge $Y$, and
$Q'$ charges for ordinary quarks and leptons. $Q_\chi$ specifies the
additional $SO(10)$ charge, whereas $Q_\psi$ is the additional
charge in the breaking of $E_6$ to $SO(10)$. Thus, $Q_\beta = Q_{\chi} \cos
\beta + Q_\psi \sin \beta$ is the general
$E_6$ charge. The charges $Q_{\chi , \psi,\eta}$  of the particular models
 correspond to $Q_\beta$  with  $\beta = 0, \frac{\pi}{2},
 -\tan^{-1} \protect\sqrt\protect{5\over 3}\ ,
$
respectively.
}
\label{charges}
\vskip 2mm
\begin{tabular}{c|cc|c|cc}
\multicolumn{1}{c|}{Fermions} & $T_{3}$
& $\sqrt {\frac {5}{3}}Y$ &
\multicolumn{1}{c|}{$Q'$} &
$2\sqrt {10}Q_{\chi}$ & \multicolumn{1}{c}{$Q_{\beta}$} \\
\hline
$ \left(\begin{array}{c}
u \\ d \\ \end{array}\right) _L $
& $ \left(\begin{array}{c}
\frac {1}{2} \\ -\frac {1}{2} \\ \end{array}\right) $
& $ \frac {1}{6} $ &
$ Q'_{q_L} $ & $ -1 $ & $-\frac{1}{2\sqrt {10}}\cos
\beta + \frac{1}{2\sqrt 6}\sin \beta$ \\
$u_L^c$ & $ 0 $ & $ -\frac {2}{3} $ &
$ Q'_{u^c_L} $ & $ -1 $ & $-\frac{1}{2\sqrt {10}}\cos \beta + \frac{1}{2\sqrt
6}\sin \beta$ \\
$d_L^c$ & $ 0 $ & $ \frac {1}{3} $ &
$ Q'_{d^c_L} $ & $ 3 $ & $\frac{3}{2\sqrt {10}}\cos
\beta + \frac{1}{2\sqrt 6}\sin \beta$ \\
$ \left(\begin{array}{c}
\nu \\ e \\ \end{array}\right) _L $
& $ \left(\begin{array}{c}
\frac {1}{2} \\ -\frac {1}{2} \\ \end{array}\right) $
& $ -\frac {1}{2} $ &
$ Q'_{\ell_L} $ & $ 3 $ & $\frac{3}{2\sqrt {10}}\cos \beta + \frac{1}{2\sqrt
6}\sin \beta$ \\
$e_L^c$ & $ 0 $ & $ 1 $ &
$ Q'_{e^c_L} $ & $ -1 $ & $-\frac{1}{2\sqrt {10}}\cos \beta + \frac{1}{2\sqrt
6}\sin \beta$ \\
\end{tabular}
\end{table}
\end{center}
\section{Model independent determination of
$Z'$ couplings at the  LHC and the  NLC}

The NLC would be able to probe in a model
independent way the following suitable combinations
of normalized gauge couplings \cite{AC}:
\begin{equation}
P_V^\ell ={{\hat g^\ell_{L} + \hat g^\ell_{R}}\over
{\hat g^\ell_{L} - \hat g^\ell_{R}}},
\ P_L^q = {{\hat g^q_{L}}\over
{\hat g^\ell_{L} - \hat g^\ell_{R}}},
\ P_R^{u,d} =  {{\hat g^{u,d}_{R}}\over
{\hat g^q_{L}}}\
 \ ,
\label{nlcc}
\end{equation}
 as well as  the ratio
\begin{equation}
 \epsilon_A=
(\hat g_{L}^\ell - \hat g_{R}^\ell)^2 {{g_2^2}\over{4\pi \alpha }}
{{s}\over{M^2_{Z'} - s}}\ \ .
\label{eps}\end{equation}

Here $\hat g_k$ are the normalized $Z'$ couplings to ordinary fermions in Eq.
(\ref{lag}),
$\hat g_k = Q'_k + \delta Y_k$  (see  Ref. \cite{LL} for  conventions and
notation).
The values of these couplings
within the general underlying $E_6$ gauge structure are given in
Table \ref{ghats}.
In Table \ref{models}
we define typical  models ($\chi , \psi , \eta ,$
$LR$, and general $LR$ models) in terms of the
parameters $\beta$ and $\delta$
 and the overall $Z'$ gauge coupling,
expressed in terms of the  electric charge $e$ and the weak angle.

On the other hand
in addition  to the mass $M_{Z'}$ the  LHC
will probe \cite{ACL} an overall strength
$g_2$  of the $Z'$ couplings as well as three
normalized couplings $\gamma_L^\ell$, $\tilde U$, and $\tilde D$, where:
\begin{equation}
\gamma_L^\ell={{(\hat{g}^\ell_{L})^2}\over
{{(\hat{g}^\ell_{L})^2+(\hat{g}^\ell_{R})^2}}},\ \
\ \  \tilde{U}={{(\hat{g}^u_{R})^2}\over {(\hat{g}^q_{L})^2}},\ \
\tilde{D}={{(\hat{g}^d_{R})^2}\over {(\hat{g}^q_{L})^2}}\ .
\label{lhcc}
\end{equation}
A fourth coupling,
\begin{equation}
\gamma_L^q={{(\hat{g}^q_{L})^2}\over{{(\hat{g}^\ell_{L})^2+(\hat{g}^\ell_{R})^2}}}
\label{lhccp}
\end{equation}
could be determined,  provided the
 $Z'$ cross section into quark pairs could be  measured with sufficient
precision. Recent studies indicate \cite{QUARK,M}
that this might be feasible. It turns out \cite{QUARK}, however, that
 for $M_{Z'}\ge 1$ TeV  and the  typical models specified in Table \ref{models}
the $Z'$  gauge couplings are too small to allow for determination of
$\gamma^q_L$ with sufficient precision at the LHC. Thus, in the
analysis the determination of  $\gamma_L^q$ is not included.

The couplings (\ref{nlcc}), probed  directly at the NLC,
 are determined with a few-fold ambiguity in
terms of the couplings  (\ref{lhcc}) and (\ref{lhccp}) probed  directly at  the
LHC.
Namely, couplings probed at the NLC (Eq. (\ref{nlcc})) are  related to the
couplings probed at the LHC (Eq.(\ref{lhcc})) in the following way:
\begin{equation}
\begin{array}{c}
{\displaystyle
P^{\ell}_V = \frac {1+2\epsilon _{\ell}
{\gamma ^\ell_L}^{\frac {1}{2}}(1-\gamma ^\ell_L)
^{\frac {1}{2}}}{2\gamma ^\ell_L - 1}}, \\
{\displaystyle
P^q_L = \frac {1}{2}\epsilon _q {\gamma ^q_L}
^{\frac {1}{2}} \left\{1+\left[\frac {1+2\epsilon _{\ell}
{\gamma ^\ell_L}^{\frac {1}{2}}(1-\gamma ^\ell_L)
^{\frac {1}{2}}}{2\gamma ^\ell_L - 1}\right]^2\right\}^{\frac {1}{2}}}, \\
{\displaystyle
P^u_R = \epsilon _u {\tilde U}^{\frac {1}{2}}}, \\
{\displaystyle
P^d_R = \epsilon _d {\tilde D}^{\frac {1}{2}}}. \\
\end{array}
\label{fewfold}
\end{equation}
where $\epsilon _{\ell}, \epsilon _q,
\epsilon _u, \epsilon _d$ can take  $\pm$ values and give rise to
a sixteen-fold sign ambiguity. The inability to probe $\gamma^q_L$ directly
implies that $P^q_L$ is not probed either.

For typical models (within the $E_6$ gauge structure), and  $M_{Z'} = 1$ TeV
the values and  expected statistical error bars for the three
 couplings (\ref{lhcc}) \cite{ACL}
and the four  couplings (\ref{nlcc})\cite{AC} are given in Table
\ref{coupl}.
 The  error bars at
the LHC  update the   analysis of Ref. \cite{ACL};
the updated numbers correspond to  the lower c.m. energy (14 TeV)
and the   (more optimistic) assumption  that the branching ratios include
$Z'$ decays  into  ordinary  three family fermions only.

In  the analysis only  statistical errors
for the observables  are included   and  error correlations for the input
parameters are neglected.
Experimental cuts
and detector acceptances  are not included  either. The
results should thus  be interpreted as a limit on how
precisely  the couplings can be determined for each model for the given
c.m. energy and the integrated luminosity of the NLC and the LHC.
 Realistic fits are expected to give larger uncertainties.
\begin{table}
\caption{$Z'$ couplings to ordinary fermions for $E_6$  models
($\beta = 0$  corresponds to  $SO(10)$ models). The
corresponding right-handed couplings are
$\hat g_R^{\psi}=-\hat g_L^{\psi^c}$.}\label{ghats}
\vskip 2mm
\begin{tabular}{c|c}
$\hat g_k$ & $Q'_k + \delta Y_k$ \\
\hline
$\hat g^q_L$
& $-\frac{1}{2\sqrt {10}}\cos \beta + \frac{1}{2\sqrt 6}\sin \beta
+ \delta \frac {1}{6} \sqrt {\frac {3}{5}}$ \\
$\hat g^{u^c}_L$
& $-\frac{1}{2\sqrt {10}}\cos \beta + \frac{1}{2\sqrt 6}\sin \beta
- \delta \frac {2}{3} \sqrt {\frac {3}{5}}$ \\
$\hat g^{d^c}_L$
& $\frac{3}{2\sqrt {10}}\cos \beta + \frac{1}{2\sqrt 6}\sin \beta
+ \delta \frac {1}{3} \sqrt {\frac {3}{5}}$ \\
$\hat g^{\ell}_L$
& $\frac{3}{2\sqrt {10}}\cos \beta + \frac{1}{2\sqrt 6}\sin \beta
- \delta \frac {1}{2} \sqrt {\frac {3}{5}}$ \\
$\hat g^{\ell^c}_L$
& $-\frac{1}{2\sqrt {10}}\cos \beta + \frac{1}{2\sqrt 6}\sin \beta
+ \delta \sqrt {\frac {3}{5}}$ \\
\end{tabular}
\end{table}
\begin{center}
\begin{table}
\caption{Parameterization of typical  $Z'$ models  within $E_6$.
The value of $\rho$
 specifies a   general $LR$ model.}\label{models}
\begin{tabular}{c|ccc}
\multicolumn{1}{c|}{Model} & $\beta $
& $\delta $ &
\multicolumn{1}{c}{$g_2$} \\
\hline
$ \chi $
& $ 0 $ & $ 0 $ & $ \sqrt {\frac {5}{3}} \frac {e}{c_W} $ \\
$ \psi $ & $ \frac {\pi}{2} $ & $ 0 $ & $ \sqrt {\frac {5}{3}} \frac {e}{c_W} $
\\   %
$ \eta $ & $ - %arc\ tan
\tan^{-1} \sqrt {\frac {5}{3}} $ & $ 0 $ & $ \sqrt {\frac {5}{3}}
\frac {e}{c_W} $ \\   %
$ LR $
& $ 0 $ & $ 0.615 $ & $ 1.382 \frac {e}{c_W} $ \\
${\rm general} \ LR $ & $ 0 $ & $ \sqrt {\frac {1}{6}}\frac {3\rho ^2-2}
{\rho ^2+1} $ & $ \sqrt {\frac {2}{5}}
\frac {\rho ^2+1}{\rho} \frac {e}{c_W} $ \\
$ \beta $ & $ \beta $ & $ 0 $ & $ \sqrt {\frac {5}{3}} \frac {e}{c_W} $ \\
\end{tabular}
\end{table}
\end{center}
\begin{center}
\begin{table}
\caption{Values  and statistical error bars for
 $\gamma^\ell_L$,  $\tilde U$,
$\tilde D$ at the  LHC  and for  $P_V^\ell$, $P_L^q$, $P_R^u$, $P_R^d$ at  the
NLC
for the $\chi$ , $\psi$ , $\eta$ , $LR$ models, with  $M_{Z'}=1$
TeV.  The error bars in parentheses are
 for the probes without polarization \protect\cite{AC}. We do not include
a possible determination of $\gamma^q_L$.}
\label{coupl}
\vskip 2mm
\begin{tabular}{c|cccc}
\multicolumn{1}{c}{} & $\chi$
& $\psi$ & $\eta$ &
\multicolumn{1}{c}{$LR$} \\
\hline
$ P_V^\ell $ & $ 2\pm 0.08(0.26) $ & $ 0\pm 0.04(1.5) $ &
$ -3\pm 0.5(1.1) $ & $ -0.15\pm 0.018(0.072) $ \\
$ P_L^q $ & $ -0.5\pm 0.04(0.10) $ & $ 0.5\pm 0.10(0.2) $ &
$ 2\pm 0.3(1.1) $ & $ -0.14\pm 0.037(0.07) $ \\
$ P_R^u $ & $ -1\pm 0.15(0.19) $ & $ -1\pm 0.11(1.2) $ &
$ -1\pm 0.15(0.24) $ & $ -6.0\pm 1.4(3.3) $ \\
$ P_R^d $ & $ 3\pm 0.24(0.51) $ & $ -1\pm 0.21(2.8) $ &
$ 0.5\pm 0.09(0.48) $ & $ 8.0\pm 1.9(4.1) $ \\
\hline
$\gamma ^\ell_L$ & $ 0.9\pm 0.016$ & $ 0.5\pm 0.02 $ &
$ 0.2\pm 0.012 $ & $ 0.36\pm 0.007 $ \\
%
%$\gamma ^q_L$ & $ 0.1  $ & $ 0.5  $ &
%$ 0.8  $ & $ 0.04  $ \\
%
$\tilde U$ & $ 1\pm 0.16 $ & $ 1\pm 0.14 $ &
$ 1\pm 0.08 $ & $ 37\pm 6.6  $ \\
$\tilde D$ & $ 9\pm 0.57 $ & $ 1\pm 0.22 $ &
$ 0.25\pm 0.16 $ & $ 65\pm 11 $ \\
\end{tabular}
\end{table}
\end{center}

\section{Reconstruction of the extended
gauge structure}

We now demonstrate how the model independent determination of the normalized
couplings in Eq. (\ref{nlcc}) or  Eq. (\ref{lhcc})  allows for  the testing
 of a particular  underlying symmetry structure and, subsequently,
how  such couplings further constrain
the  symmetry breaking pattern.   We demonstrate the procedure for
models based on $E_6$  gauge symmetry.

As the first step in establishing the  underlying $E_6$ symmetry structure, the
 normalized couplings should  be compatible  with the discrete
constraints on the couplings  of the type (\ref{const}), which,  in terms of
the
$\hat g$  couplings, are:
\begin{equation}
\begin{array}{c}
{\displaystyle
2 \hat g^q_{L} + \hat g^u_{R} + \hat g^\ell_{R} = 0}, \\
{\displaystyle
\hat g^q_{L} - \hat g^d_{R} - \hat g^\ell_{L} + \hat g^\ell_{R} = 0}, \\
{\displaystyle
3 \hat g^q_{L} + \hat g^\ell_{L} = 0}. \\
\end{array}
\label{coupcons}\end{equation}
Only the first two equalities hold for general  $E_6$ models. The third
holds in the special case of $SO(10)$.
At the NLC these constraints, expressed in terms of (\ref{nlcc}), are of
the form:
\begin{equation}
\begin{array}{c}
{\displaystyle
C_1 \equiv 2 P^q_L (2 + P^u_R) + P^{\ell}_V - 1 = 0 \ ,} \\
{\displaystyle
C_2 \equiv P^q_L (1 - P^d_R) - 1 = 0\ ,}\\
{\displaystyle
C_3 \equiv 6 P^q_L + P^{\ell}_V + 1 = 0\ \ .}  \\
\end{array}
\label{nlcco}
\end{equation}
The  underlying gauge structure is  therefore  determined by the above
constraints; the symmetry
group corresponds to $SO(10)$
if  the
measured normalized couplings are compatible with all
three constraints (\ref{nlcco}), while the symmetry group
is  $E_6$ if  the
measured  couplings  are compatible with the first
two  only.

The next step is to address
the nature of the symmetry breaking pattern
within the underlying symmetry structure.

The  normalized couplings probed
at the NLC (see Eq. (\ref{nlcc}))
in turn determine $\tan \beta $ and $\delta $, parameterizing the most general
symmetry breaking pattern within the $E_6$ group (recall $\beta=0$
corresponds to the $SO(10)$ group).
$\beta$ and $\delta$ can be expressed in terms of the normalized couplings
by:
\begin{equation}
\begin{array}{c}
{\displaystyle
\tan \beta = \sqrt {15} \frac {P^q_L(1 - P^u_R) + 1}
{P^q_L(1 + 3 P^u_R - 4 P^d_R) - 1 + 2
P^\ell_V }}, \\
{\displaystyle  \frac {\delta}{\cos \beta} = 2 \sqrt 6 \frac {P^q_L(1 + P^u_R)}
{P^q_L(1 + 3 P^u_R - 4 P^d_R) - 1 + 2 P^\ell_V }\ \ .} \\
\end{array}
\label{nlcdb}
\end{equation}

Determination of the symmetry breaking pattern within  the underlying gauge
structure can be thought as replacing the phenomenological
variables in (\ref{nlcc}) by the four independent variables
$C_1, C_2, \beta , \delta $ in (\ref{nlcco}) and
(\ref{nlcdb}). ($C_3$ is a function of $C_1$ and $\beta $.)
$\beta $ and $\delta $ are interpreted as
determining the pattern of symmetry breaking within $E_6$
for $C_1 = C_2 = 0$ only.
\begin{itemize}
\item
One can determine the parameters in (\ref{nlcc}) and their
correlation matrix from the NLC probes, and from these
determine $C_1$, $C_2$, $\beta$ ,  and $\delta $ ({\it step one}). If $C_1$ and
$C_2$ are compatible with zero, then one  assumes
$C_1 = C_2 = 0$ ({\it step two}), {\it i.e.},  one  assumes $E_6$  to be  the
underlying symmetry. In this case only two normalized
couplings in (\ref{nlcc}) are independent, thus
in turn  implying that  $\beta $ and $\delta $  can be determined with better
precision.
\item
Equivalently, one can rewrite  the NLC probes as functions
of $C_1, C_2, \beta , \delta $ and fit them directly to
experimental data ({\it step one}). If $C_1$ and
$C_2$ are compatible with zero, then one can fix them to be zero ({\it step
two}), {\it i.e.},  one is assuming $E_6$ as the underlying symmetry,  and
then fit $\beta $ and $\delta $
with higher precision.
\end{itemize}
We will use the  second approach
in our numerical examples  below.

Before proceeding with the numerical analysis,
 we will introduce   new convenient  combinations of couplings,
which in the case of $E_6$ symmetry are set to zero. The
 choice of variables $C_1, C_2, \beta$, and $\delta $ is
suggested by the $E_6$ parameterization  of charges.
There is an equivalent choice of parameters:
\begin{equation}
S_1 = C_2, \
S_2 = 2 (2 + P^u_R) + ( 1 - P^d_R ) ( P^\ell_V - 1 ), \
\beta , \
\delta .
\label{nlcnv}
\end{equation}
$S_2$ is a linear combination of $C_1$ and $C_2$, which does not depend on
$P^q_L$.
 Thus,    $C_1 = C_2 = 0$  if and only if   $S_1 = S_2 = 0$, {\it i.e.},
 the latter  set of constraints  uniquely fix $E_6$ as the
underlying symmetry as well.
In the following we shall  use  variables (\ref{nlcnv})
 $S_1, S_2, \beta$ ,  and $\delta $, because they  are better adapted
to the analogous analysis  at the LHC, and thus
allow for  an easy comparison of the NLC and the LHC potentials.

The analogous analysis at LHC requires one to rewrite
(\ref{nlcco}), (\ref{nlcdb}), (\ref{nlcnv})
in terms of (\ref{lhcc}) and (\ref{lhccp}).
This can be done using  the relation (\ref{fewfold}).
The sign ambiguity due to  different sign assignments for
$\epsilon _{\ell}, \epsilon _q,
\epsilon _u,$, and $ \epsilon _d$  give rise to
a sixteen-fold sign ambiguity due to the fact that
at the LHC only the
magnitude of the corresponding  couplings is determined.
In general, $C_1 = C_2 = 0$ ($S_1 = S_2 = 0$) can be
satisfied only for a specific choice of the
corresponding signs.\footnote{Note, however, that for  special values of
$\delta$ and/or $\beta$ parameters  a few-fold sign ambiguity is not removed.
}

\subsection{Determination of the symmetry  breaking structure at the NLC}

As stated above we fit the expected NLC observables
for $S_1, S_2, \beta , \delta$ and
$\epsilon_A$.
In Table \ref{sigma} the values and  1 $\sigma$ uncertainties for
these quantities   are given for the models specified in
Table \ref{models} and $M_{Z'}=1$ TeV.  The parameters
$\beta$ and $\delta$ can
be determined  with uncertainties  between $ 0.02$ and $0.06$
 for a large class of typical  models.
 The large uncertainties for the $\eta$ model are
primarily due to the small value of  $\epsilon_A$.
\begin{figure}[p]
\iffiginc
\vspace{-0.5in}
\psfig{figure=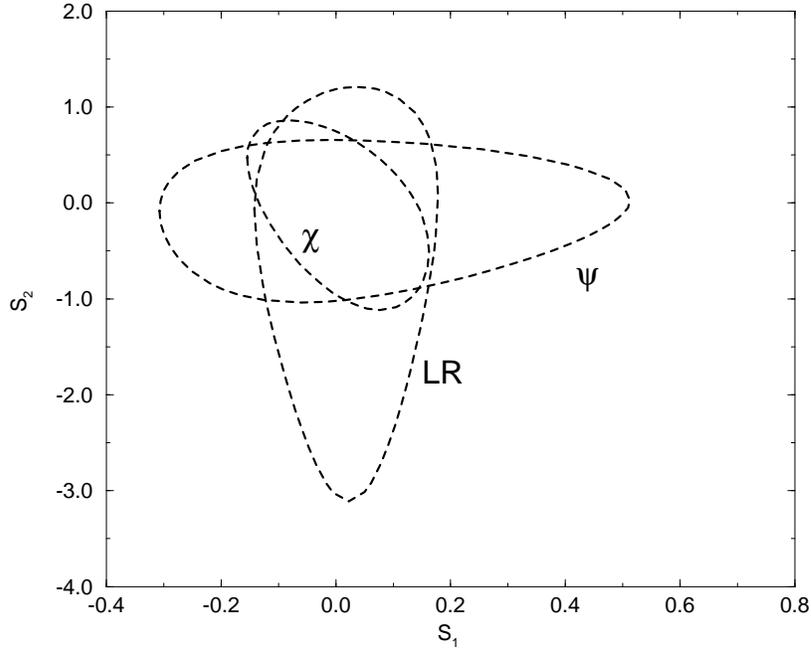,height=4in,angle=-90}\hfill
\else
\fi
\caption{90\%  confidence level   ($\Delta \chi^2=4.6$) contours  for
 $S_1$  {\it vs.} $S_2$ (constraints defined in Eq.
(\protect\ref{nlcnv}) for the  typical models
%(defined in Table \protect\ref{models})
at the  NLC.   $M_{Z'}=1$ TeV.
Only statistical error bars for the probes are  used.}
\label{nlcs1s2}\end{figure}
\begin{figure}[p]
\iffiginc
\vspace{-0.5in}
\psfig{figure=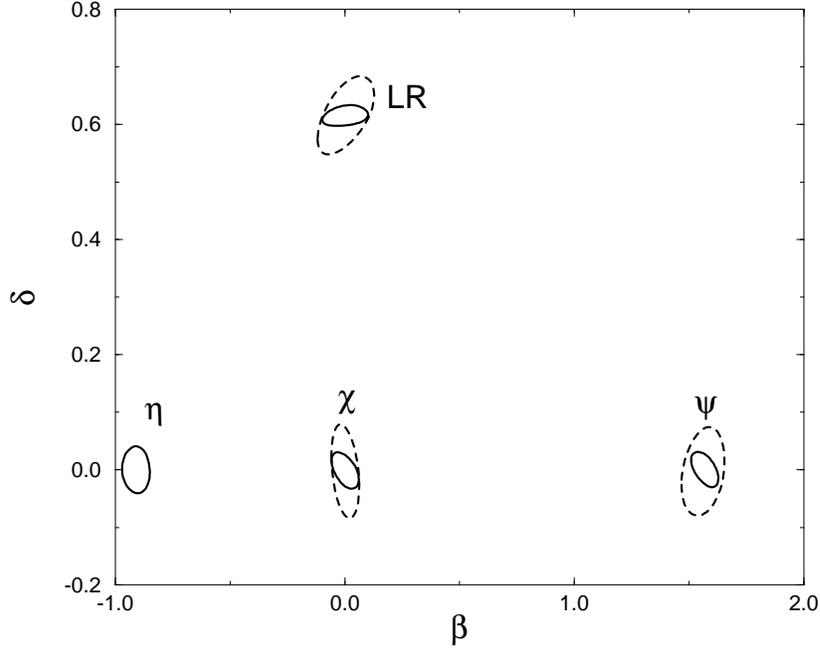,height=4in,angle=-90}\hfill
\else
\fi
\caption{ 90\%
confidence level   ($\Delta \chi^2=4.6$) contours
(dashed lines) for  $\beta$ {\it vs.} $\delta$
for the  typical models
at the NLC. $M_{Z'}=1$ TeV. The solid lines correspond to the
assumption that the constraints $S_1 = S_2 = 0$ are satisfied.
Only statistical error bars for the probes are  used.}
\label{nlcbvd}
\end{figure}
In Figure \ref{nlcs1s2} we plot  the $ 90\%$ confidence level
($\Delta \chi ^2 =4.6 $)
contours (dashed lines) for two constraints
($S_1$ and $S_2$ as defined in (\ref{nlcnv}))
 for the  models  specified in Table
\ref{models} and $M_{Z'}=1$ TeV. For the $\eta$ model  the uncertainties  are
slightly
 too large  to allow for the corresponding
 90\% confidence level  plot.
In Figure  \ref{nlcbvd} (dashed lines)
we plot the $ 90\%$ confidence level ($\Delta \chi ^2 =4.6 $)
contours in the $\beta$ {\it vs.} $\delta $
 plane for the typical models and $M_{Z'}=1$ TeV.
\begin{center}
\begin{table}
\caption{Values  and  1 $\sigma$  statistical  error bars for
$S_1, S_2, \beta , \delta$,  and  $\epsilon_A$ at the NLC
 for the typical models.
 $M_{Z'}=1$ TeV.
The error bars in parentheses are
determined by setting
  $S_1=S_2=0$.}
\vskip 2mm
\begin{tabular}{c|cccc}
\multicolumn{1}{c}{} & $\chi$
& $\psi$ & $\eta$ &
\multicolumn{1}{c}{$LR$} \\
\hline
$S_1$ & $ 0\pm 0.074 $ & $0\pm 0.18 $ &
$0\pm 0.22$ & $ 0\pm 0.074 $ \\
$S_2$ & $ 0\pm0.45$ & $0\pm0.37$ &
$ 0\pm0.51 $ & $0\pm 0.76 $ \\
$\beta$ & $ 0\pm 0.028 (.028)$ & $1.57\pm 0.043 (0.027) $ &
$ -0.912\pm 0.038 (0.028)$ & $ 0\pm 0.058 (0.047)$ \\
$\delta$ & $ 0\pm 0.038 (0.015) $ & $0\pm0.035 (0.014) $ &
$ 0\pm 0.059 (0.019)$ & $0.615\pm 0.032 (0.008)$ \\
$\epsilon_A$ & $0.071\pm 0.005 (0.005)$ & $ 0.121\pm 0.017 (0.010) $ &
$0.012\pm0.003 (0.003) $ & $0.255\pm 0.016 (0.009)$ \\
\end{tabular}
\label{sigma}
\end{table}
\end{center}

As a second step we fix  $S_1, S_2$
to zero, fitting the values of $\beta$, $\delta$ and
$\epsilon_A$. Clearly, the uncertainties are reduced.  In Table \ref{sigma}
the corresponding 1 $\sigma$ uncertainties are given in  parentheses. In
this case even the parameters of the  $\eta$  model can be determined with a
sufficient accuracy. The uncertainties are now reduced by a factor
of 1 to 4 compared to the step one analysis.
In Figure 2 the $ 90\%$ confidence level
($\Delta \chi ^2 =4.6 $)
contours in the $\beta$ {\it vs.}   $\delta $ are
plotted with solid lines.

Provided the mass $M_{Z'}$ is known by
the time the NLC is turned on, the  determination of the  $\epsilon_A$
parameter would yield  additional useful information on the overall strength
$g_2$ and thus on the symmetry breaking patterns\cite{RR}.

\subsection{Determination of the symmetry breaking structure at the LHC}

At the LHC
primarily three  couplings (\ref{lhcc})
 would be probed for a large class of
models\footnote{The LHC would also determine the $Z'$ mass. It
 would furthermore establish whether there is a corresponding $W'$, as
 expected in $LR$ models, and the ratio $M_{W'}/M_{Z'}$, which probes
 the $LR$-breaking mechanism \cite{lrbreaking}.}.
 Namely,
for $M_{Z'}=1$ TeV and typical models  specified in Table \ref{models}
 $\gamma ^{\ell}_q (P^q_L)$ cannot be
determined with sufficient precision. Consequently,
one can only determine the three  combinations of
$S_2, \beta$, and $ \delta $. We therefore set $S_1 = 0$,
thus expressing the undetermined coupling $P^q_L$ in terms of the three
left-over  couplings $P^\ell_V$, $P^u_R$, and  $P^d_R$.  This constraint also
removes one sign ambiguity ($\epsilon_q$).
Then, we  fit for $S_2, \beta $,  and $\delta $.
Thus, at the LHC  one  has to  assume that one  ($S_1=0$) of the two $E_6$
discrete constraints  on the gauge couplings is already satisfied in order to
further test whether or not the  second discrete constrai $S_2=0$ is
satisfied and then ultimately determine the values of $\beta$ and $\delta$.

\begin{figure}
\iffiginc
\vspace{-0.5in}
\psfig{figure=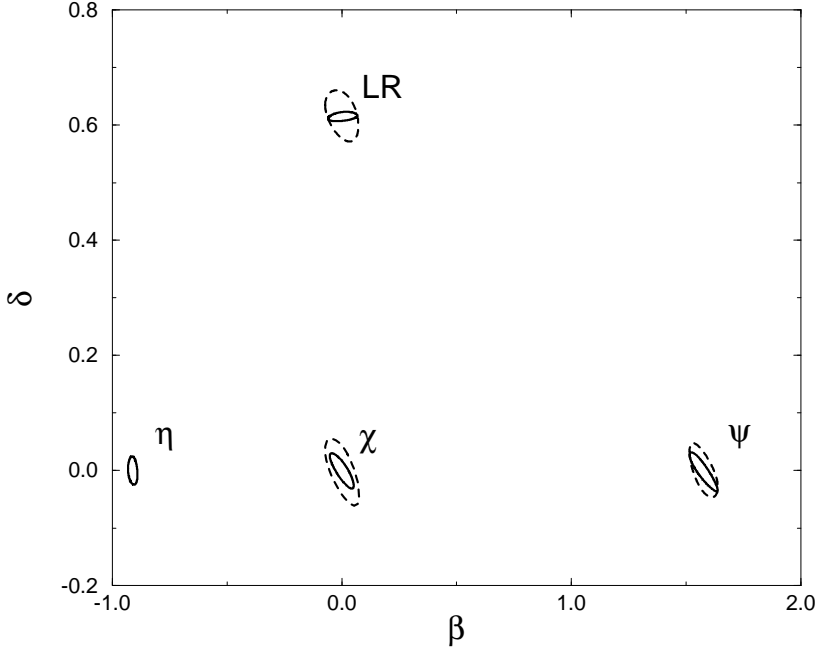,height=4in,angle=-90}\hfill
\else
\fi
\caption{ 90\%
confidence level   ($\Delta \chi^2=4.6$) contours  for  $\beta$ {\it vs.}
$\delta$
for the  typical models
at the LHC.   $M_{Z'}=1$ TeV.
Only statistical error bars for the probes are  used. Dashed lines
are determined by fixing $S_1=0$, while the solid ones correspond to setting
$S_2=0$ as well.}
\label{lhcbvd}
\end{figure}

In Table \ref{sigmal} the values and  expected 1 $\sigma$ uncertainties for
these quantities   are given for the models specified in
Table \ref{models} and $M_{Z'}=1$ TeV.
In general, there are eight disjoint
regions  corresponding
to the different sign assignments for $\epsilon _u, \epsilon _d,
\epsilon _{\ell}$. We only quote the results for the region  in the vicinity
of $S_2 = 0$.\footnote{In general, the region in the vicinity of $S_2=0$
 removes the few-fold sign ambiguity, thus uniquely fixing  the value
of $\beta$ and $\delta$. However, for special models,
such a  region  can be fitted with more than one
 value of $\beta$ and $\delta$. {\it E.g.}, the couplings corresponding to
the $\chi$, and $\eta$ models, can also be  fitted to the models
with $\delta=0$,
 $\beta=\tan^{-1}(3\sqrt{3/5})$,  and $\delta=0$,
$\beta=-\tan^{-1}(7\sqrt{3/5})$, respectively.
The  model with $\delta=0$, $\beta=\tan^{-1}(\sqrt{3/5})$ is an alternative
 of the $\psi$ model. In this case
constraints $C_1=C_2=0$ (or $S_1=S_2=0$) are formally undetermined because they
are obtained by dividing  the first two constraints in (\ref{coupcons}) by
$\hat
g_L^\ell-\hat g_R^\ell=0$.  However, the first two constraints  in
(\ref{coupcons})
 are satisfied.  In Figure \ref{lhcbvd}  we do
not display any of the alternative  regions.
 }
The parameters
$\beta$ and $\delta$ can
be determined  with uncertainties  between 0.02 and 0.04
 for a large class of typical  models.

In Figure \ref{lhcbvd}
we plot the $ 90\%$ confidence level ($\Delta \chi ^2 =4.6 $)
contours  (dashed lines) in the $\beta$ {\it vs.}  $\delta $ plane for the
typical
models  specified in Table
\ref{models} with $M_{Z'}=1$ TeV, and for the region  in the vicinity
of  $S_2 = 0$.
\begin{center}
\begin{table}
\caption{Values  and  1 $\sigma$  statistical  error bars for
$S_2, \beta , \delta$ for the typical models at the LHC. $S_1=0$ and
 $M_{Z'}=1$ TeV.
The error bars in parentheses are
determined by setting $S_2=0$.}
\vskip 2mm
\begin{tabular}{c|cccc}
\multicolumn{1}{c}{} & $\chi$
& $\psi$ & $\eta$ &
\multicolumn{1}{c}{$LR$} \\
\hline
$S_2$ & $ 0\pm0.40$ & $0\pm0.16$ &
$ 0\pm0.71 $ & $0\pm 0.54 $ \\
$\beta$ & $ 0\pm 0.024 (0.024)$ & $1.57\pm 0.028 (0.028) $ &
$ -0.912\pm 0.010 (0.010)$ & $ 0\pm 0.034(0.029)$ \\
$\delta$ & $ 0\pm 0.019 (0.014) $ & $0\pm0.022 (0.016) $ &
$ 0\pm 0.016 (0.011)$ & $0.615\pm 0.021 (0.004)$ \\
\end{tabular}
\label{sigmal}
\end{table}
\end{center}
As a second step, we set   $S_2$  to zero as well.
 %In general, this constraint
%would remove  the eight-fold  ambiguity.
In this case  the fitted values  for
 $\beta$ and $\delta$  have
 smaller uncertainties. In Table \ref{sigmal}
the corresponding 1 $\sigma$ uncertainties are given
in  parentheses.
The uncertainties are now reduced by a factor
of 1 to 5 compared to the step one analysis.
In Figure \ref{lhcbvd} the $ 90\%$ confidence level
($\Delta \chi ^2 =4.6 $)
contours  in  $\beta$ {\it vs.}   $\delta $ are
plotted with solid lines.
The  parameters  can be determined
with a precision of about a few percent.
For the typical models  the  uncertainties are
 smaller than those at the NLC (see Figure \ref{nlcbvd}).

Although  the overall strength of the
$Z'$ gauge couplings  does  not enter in the above analysis
(except for the estimate of typical statistical error), the overall
strength of the interactions $g_2^2((\hat{g}^\ell_{L})^2+
(\hat{g}^\ell_{R})^2)$ can also be measured at the LHC, by measuring  the
cross-section  in the main production channel and  the  total
width. Determination of this  coupling  can
be  used  to constrain  further the symmetry breaking pattern\cite{RR}.

\section{Conclusions}

We have demonstrated  how the  model independent determination of gauge
couplings of a possible $Z'$ at  future colliders would allow
one to gain information about the  nature of
the extended gauge structure associated with the $Z'$.

As a prime case we study $E_6$ [$SO(10)$]  as the underlying
gauge symmetry. In the case of $E_6$ [$SO(10)$]  $Z'$ gauge couplings
 have to satisfy  two [three] discrete constraints.
Provided such constraints are  satisfied,
the parameters
$\tan \beta$ [$\beta=0$  for $SO(10)$] and $\delta$, which
characterize the effects of  the  symmetry
pattern within $E_6$ [$SO(10)$] on the $Z'$ couplings, can
be determined.

For  $M_{Z'}\sim 1$
--2 TeV and typical models
 the statistical uncertainties  for    parameters $\beta$
and $\delta$
are in the 0.01--0.04 range, once the  corresponding
discrete constraints are
fixed.  We included only statistical uncertainties
for the probes.
Realistic fits, which include experimental cuts
and detector acceptances,  are expected to give larger uncertainties.

\acknowledgments

The work was supported  in part
by CICYT under contract AEN94-0936 (F.del A.), the European Union under
contract CHRX-CT92-0004 (F. del A.), the Junta de Andaluc\'\i a  (F. del A.)
 and the  U.S. DOE  Grant No. DOE-EY-76-02-3071 (M. C. and P.L.).

\end{document}